# A density functional theory investigation of charge mobility in titanyl-phthalocyanines and their tailored peripherally substituted complexes

Jeffrey R. De Lile, and Su Zhou[*]

School of Automotive Studies, Tongji University,

4800 Cao'an Road, 201804, Shanghai,

P.R. China

**Corresponding author, E-mail address: suzhou@tongji.edu.cn**

**Abstract**

Titanyl-phthalocyanines catalytic ability towards oxygen reduction is demonstrated in experimental literature. Our recent theoretical simulations revealed electronic structure origin of catalytic ability in peripherally and axially substituted triplet and singlet titanyl-phthalocyanines. However, the origin of high electron transfer ability to spontaneously reduce peroxide in chlorine substituted singlet complex and triplet state Ti(II)Pc complexes remain elusive. Thus, we performed density functional theory calculations to study Ti(IV)Pc and their tailored peripheral substituted complexes as representative compounds of titanyl-phthalocyanines for charge mobilities, reorganization energies and electronic couplings. In addition, oxo(phthalocyaninato)titanium(IV) (TiOPc) convex and concave compounds were investigated to benchmark the method. Based on the results, higher charge mobility and reorganization energy associated with electron transfer of TiOPc are predicted with reasonable accuracy. Reorganization energies of triplet state Ti(II)Pc and their tailored peripheral substituted complexes are compared with Ti(IV)Pc singlet complexes in order to understand the charge mobility. Chlorine substituted complex demonstrates higher electron hopping rate due to higher electronic coupling in comparison to other halogens. Similarly, weak electron-donating methyl group increases the electron transport rate. Moreover, halogen substituted Ti(II)Pc complexes elucidate lower reorganization energies. The lowest reorganization energy is predicted for chlorine substituted Ti(II)Pc complex, which is 0.09 eV. Therefore, higher electronic coupling and lower reorganization energies can be considered as the origin of higher electron transfer ability of these complexes. Furthermore, increase electron hopping rate due to weak electron-donating substituents provide a method to produce efficient n-channel organic field-effect transistors with higher electron mobility.

**Introduction**

The energy landscape of the future is determined by the high surface area conductive materials. Microporous conductive polymers, ceramics, chalcogenides and activated carbons have been applied in electrocatalysis, sensing, separation methods, supercapacitors and high temperature fuel cells [1]. Metallophthalocyanines (MPcs) are one class of materials that have a high potential in this area. The MPcs are large macrocyclic molecular complexes with extended π system. They are applied in various technologies. The MPcs have been used as pigments for blue and green colors, electro-chromic materials, photoconductors, organic solar cells, photo-dynamic tumor therapy, organic light emitting diodes, organic field effect transistors, dye sensitized solar cells, LCD displays, CD technology and catalysis [2].

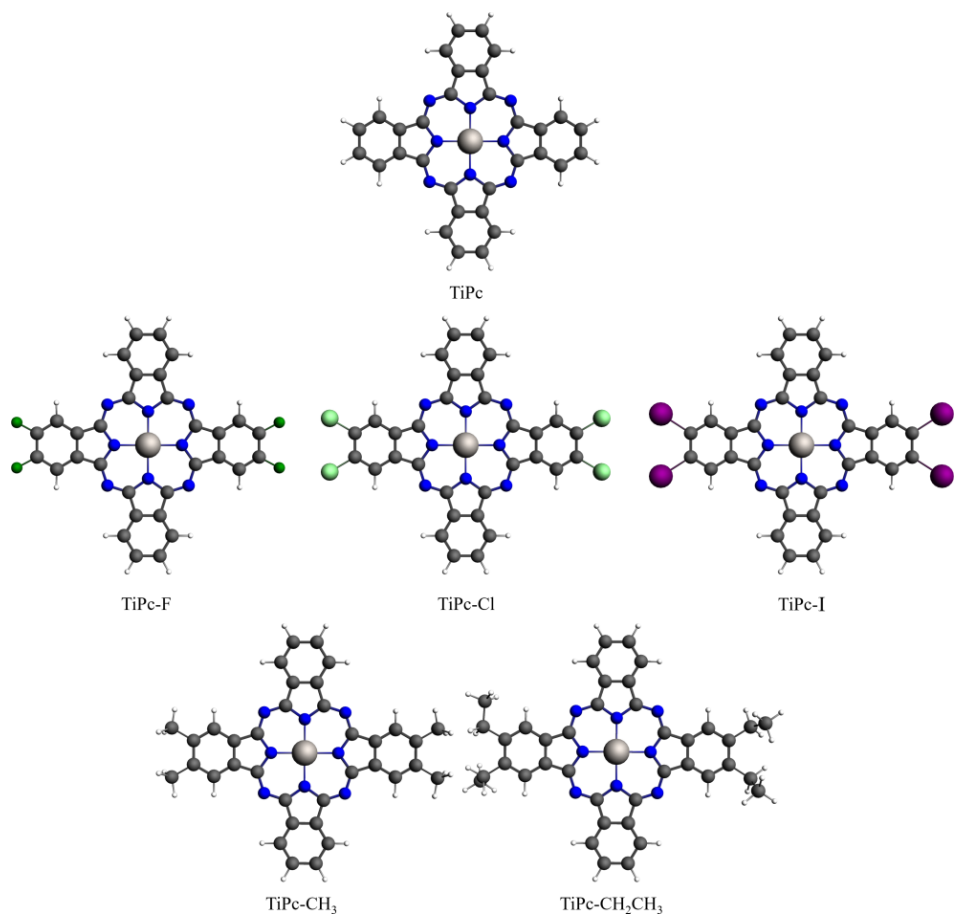

**Fig- 1** Titanyl-phthalocyanine complexes investigated in this study. Here chlorine is light green, iodine purple, fluorine dark green, titanium silver, nitrogen blue, carbon grey and hydrogen white in color. Abbreviated names are listed under the molecular complex .

Besides conductivity, their catalytic ability towards oxygen reduction reaction (ORR) have attracted much attention recently. Among large array of MPcs, iron-phthalocyanine (FePc) is proposed as the best available phthalocyanine based catalysts for ORR in PEMFCs [3]. However, our recent theoretical work suggested that triplet state titanyl-phthalocyanine (Ti(II)Pc) and their peripherally modified complexes as well as peripherally halogen substituted singlet [Ti(IV)Pc]$^{2+}$ perform oxygen reduction much better [4]. Figure-1 illustrates the model structures of the titanyl-phthalocyanines that have been used in this study. Figure-2 presents reduction of activation barrier for peroxide dissociation due to fluorine substitution. In fact, it is predicted to reduce hydrogen peroxide, the rate limiting step of the oxygen reduction on MPcs, spontaneously without any activation barrier. Therefore, those complexes increase the kinetics of the ORR process dramatically. This ability particularly found in chlorine substituted singlet [Ti(IV)Pc]$^{2+}$ and all the triplet Ti(II)Pc complexes [4]. In addition, we observed that spontaneous peroxide reduction directly related to higher charge transfer ability of those complexes. However, the origin of higher charge transfer ability in these compounds remain elusive. Therefore, density functional based theoretical simulation has been employed to study charge mobility between molecules. The electronic coupling between molecules and reorganization energies are the only simulated parameters used to determine charge hopping rate using Markus rate theory [5]. However, [Ti(IV)Pc]$^{2+}$ and its peripherally modified complexes have 2+ charge. Therefore, it is not possible to calculate reorganization energies to determine charge transfer ability under the present theoretical framework. Because it is always calculated between neutral state and the plus one (or minus one) charge state of the complexes. On the other hand, Ti(II)Pc triplet state complexes cannot be calculated for electronic coupling under DFT method to obtain reasonable results. Thus we used (Ti$^{4+}$Pc$^{4-}$) Ti(IV)Pc neutral titanyl-phthalocyanine as a representative case to understand charge mobility in these complexes. The results obtained is directly compared with reorganization energies of Ti(II)Pc complexes to gain insight into the charge transport phenomenon.

Therefore, in this letter we presented for the first time reorganization energies associated with electron transfer, electronic coupling and Markus hopping rates of Ti(IV)Pc and their tailored peripheral substituted complexes. In addition, reorganization energies of Ti(II)Pc is presented and discussed the origin of higher charge transfer ability based on the available results. Moreover, TiOPc convex and concave structures are calculated for Markus hopping rates and compared with the available literature to bench mark the method.

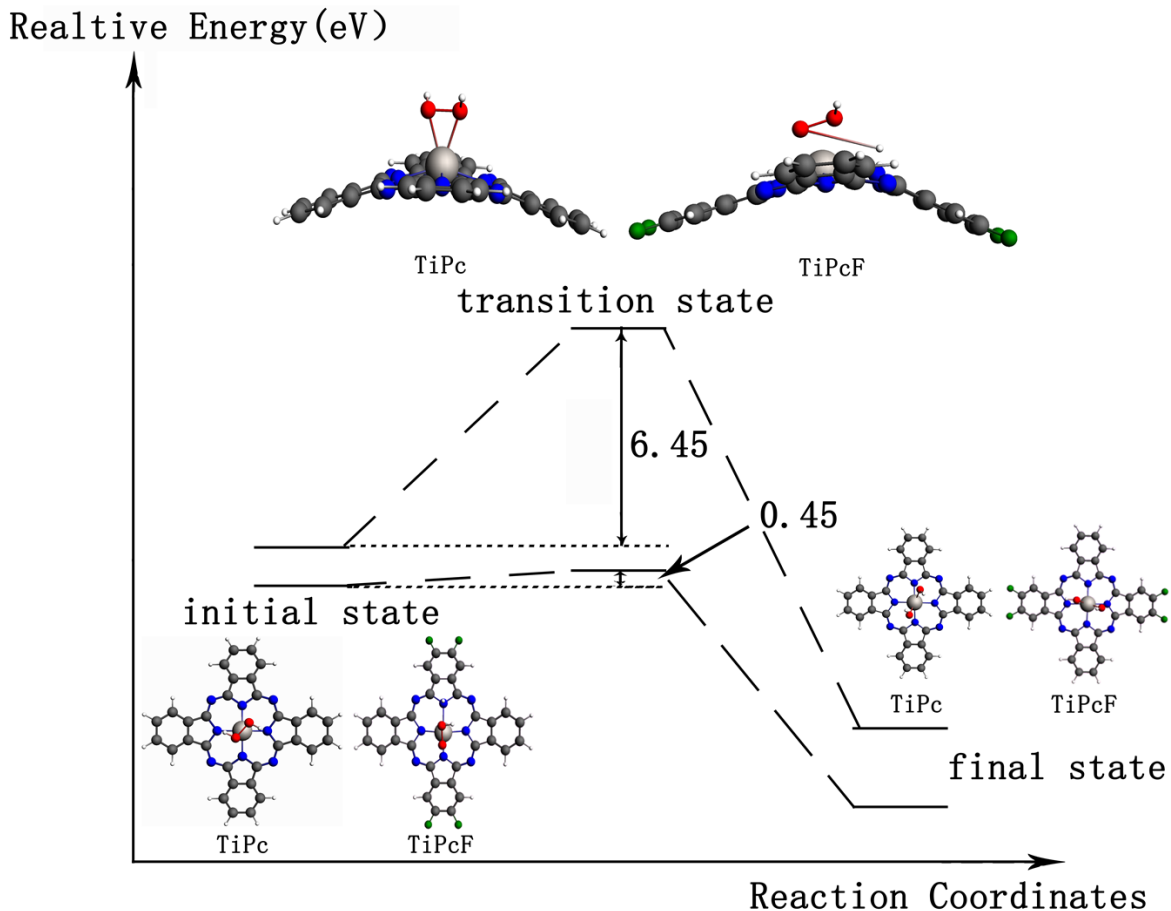

**Fig- 2** Initial, final and transition states of [Ti(IV)Pc]$^{2+}$ and [Ti(IV)Pc-4F]$^{2+}$ complexes. Note the dramatic reduction of activation energy barrier from fluorine substituted complex from 6.45 to 0.45 eV. Here oxygen is red, titanium silver, fluorine green, nitrogen blue, carbon grey and hydrogen white in color. Adopted from ref [4].

**Theoretical calculation**

The electronic coupling between the molecules and reorganization energies associated with electron transfer were calculated using ADF software package [6]. For the electronic coupling calculations TZP basis set with PW91 [7] generalized gradient approximated exchange-correlation functional have been employed. However, carbon, nitrogen, hydrogen, oxygen and fluorine were calculated using DZP basis set in both electronic coupling and reorganization energy calculations. Only titanium, chlorine and iodine were simulated with TZP basis set. All electron basis sets were used for all the calculations (no frozen core approximation). The exchange-and correlation energies were represented by hybrid level B3LYP [8] functional for reorganization energy calculations. The scalar ZORA relativistic [9] method has been used to study the relativistic effects. The Markus-Hush rate equation of electron hopping was used to calculate electron and hole transport rates [5]. In this letter substituted and unsubstituted Ti(IV)Pc molecules were kept at 3.118 Å distance irrespective of their crystallographic parameters to compare them with TiOPc electronic coupling values. The TiOPc single crystal parameters were obtained from the reference [10]. Due to non-planar nature of TiOPc molecule, both concave and convex structures were analyzed for charge transportation rates.

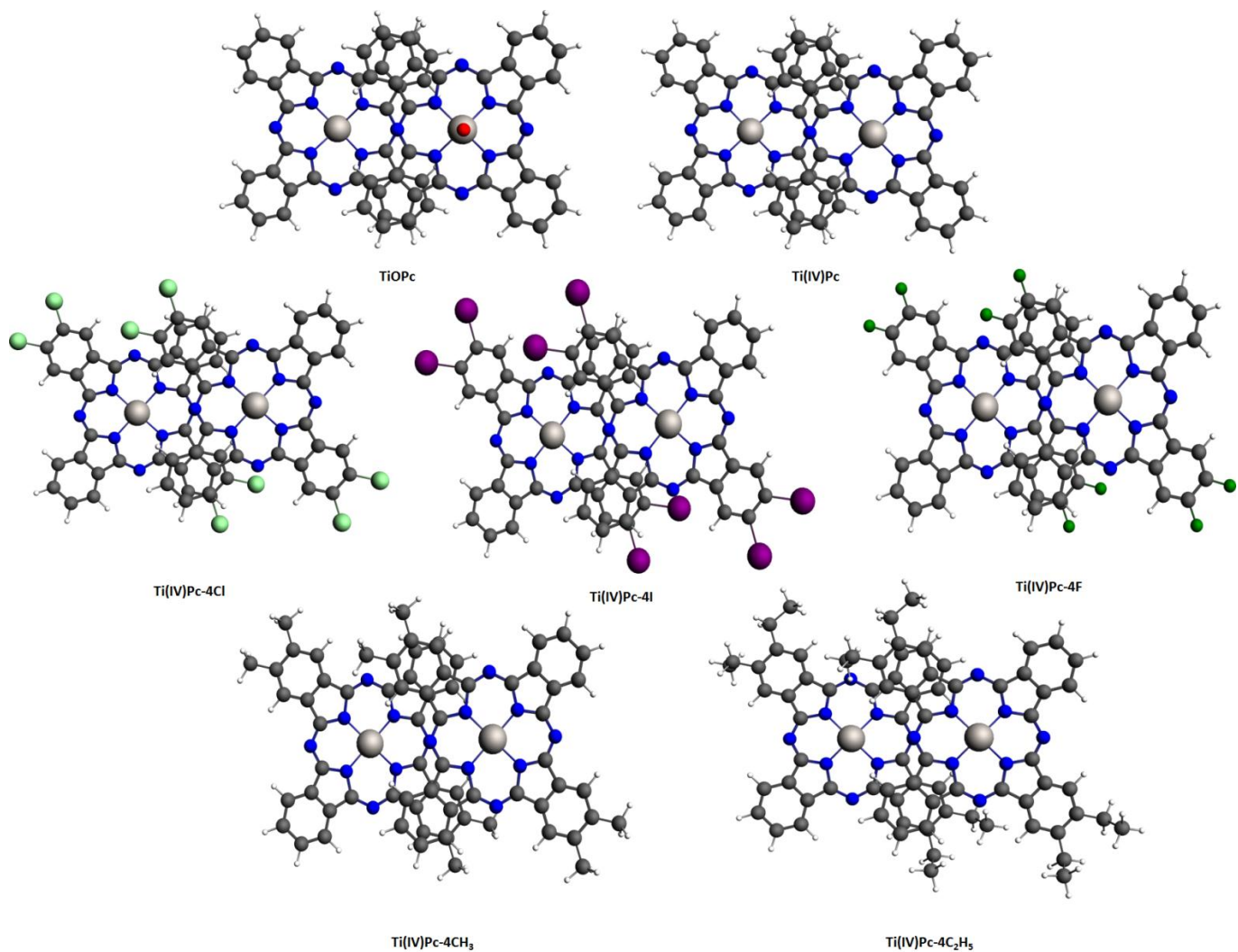

**Fig- 3** The arrangement of molecular complexes for electronic coupling calculations. The distance between two molecules is 3.118 Å and that is defined by the plane consisting of aza-nitrogen atoms (nitrogen atoms bonded only to two carbons). The overlapping aza-nitrogen and aza-benzyl rings can clearly be seen. Here nitrogen is blue, titanium silver, carbon grey, hydrogen white, oxygen red, chlorine light green, fluorine dark green, Iodine purple in color

In order to determine reorganization energy, four geometry optimization calculations were performed on a molecule to find the neutral ground state of the molecule ($E_{neutral}$), excited state (anion) energy of the molecule on its ground state geometry ($E_{anion}$(neutral geometry)), excited state geometry ($E_{anion}$) and neutral state energy of the molecule on its excited state geometry ($E_{neutral}$(anion geometry)). The equation (1) has been used to calculate reorganization energy ($\lambda$) and the electronic coupling (V) is calculated using equation (2) as described in the ADF user guide.

$$\lambda = (E_{anion}(\text{neutral geometry}) - E_{neutral}) + (E_{neutral}(\text{anion geometry}) - E_{anion}) \quad (1)$$

$$V = \left(\frac{J - S(\varepsilon 1 + \varepsilon 2)/2}{1 - S^2}\right) \quad (2)$$

Where V is electronic coupling, J is charge transfer integral between highest occupied molecular orbitals (HOMOs) of the two molecules in case of hole transfer or lowest unoccupied molecular orbitals (LUMOs) in case of electron transfer. S is overlap integral between HOMOs of two molecules (or LUMOs) and ε1 & ε2 are site energy of HOMO (LUMO) of molecule one and molecule two respectively. Thus based on these two theoretically calculated results, Markus rate for electron and hole hopping can be determined.

$$k = \frac{V^2}{\hbar}\sqrt{\frac{\pi}{\lambda k_B T}} \exp\left(-\frac{\lambda}{4 k_B T}\right) \quad (3)$$

The adiabatic Markus-Hush equation (equation 3) has been used to calculate hole and electron hopping rates. Where V is electronic coupling between the molecules, $\lambda$ is the reorganization energy, $\hbar$ is the reduced Plank's constant, $k_B$ is the Boltzmann constant and T is the temperature in kelvin. However, temperature was usually taken as the room temperature of 300 K during the calculations. Moreover, the electronic couplings and reorganization energies were used in units of joules in Markus-Hush rate equation.

**Results and discussion**

The Markus rate values for the electron and hole hopping between the molecules of TiOPc, Ti(IV)Pc and their tailored peripheral substituted complexes are tabulated in table-1. In addition, reorganization energies associated with electron transfer and electronic coupling values are also tabulated. Reorganization energies calculated for Ti(II)Pc and their peripherally modified complexes are presented in table-2. Figure-3 illustrates the arrangement of the titanyl-phthalocyanine pairs for the electronic coupling calculations. The molecules are separated at a 3.118 Å distance corresponding to crystallographic structure of convex pair. The distance is measured from the plane defined by the aza-nitrogen atoms. The results correctly predicted TiOPc and Ti(IV)Pc, which are p-type semi-conductors, have higher hole hopping rates than that of electrons [10,11]. In addition, reorganization energy associated with electron transfer of TiOPc is reported to be 0.180 eV [10], is very close to our prediction 0.16 eV. Therefore, our calculations agreed well with the available literature. The concave and convex molecular pairs of TiOPc illustrates in Figure-4.

**Table 1** DFT calculated parameters and hopping rates. All the energy values are in eV and the hopping rate has units per second ($S^{-1}$). TiOPc couplings in ref [10] are tabulated for comparison.

| Complexes | Electronic Coupling (V) of holes, eV | Electronic Coupling (V) of electrons, eV | Reorganization Energy ($\lambda$), eV | Markus Hopping rates for hole, $s^{-1}$ | Markus Hopping rates for electrons, $s^{-1}$ | Ref 10 hole, eV | Ref 10 electron, eV |
|---|---|---|---|---|---|---|---|
| TiOPc concave | -0.280 | 0.008 | 0.16 | $6.98E^{+14}$ | $1.21E^{+11}$ | -0.049 | -0.037 |
| TiOPc convex | -0.148 | 0.011 | 0.16 | $1.97E^{+14}$ | $1.98E^{+11}$ | -0.136 | -0.033 |
| Ti(IV)Pc | -0.198 | -0.003 | 0.36 | $3.35E^{+13}$ | $8.09E^{+9}$ | | |
| Ti(IV)Pc-4F | 0.019 | 0.004 | 0.62 | $1.97E^{+10}$ | $9.73E^{+8}$ | | |
| Ti(IV)Pc-4Cl | 0.010 | 0.008 | 0.68 | $2.67E^{+9}$ | $2.00E^{+9}$ | | |
| Ti(IV)Pc-4I | 0.012 | 0.004 | 0.72 | $2.92E^{+9}$ | $2.58E^{+8}$ | | |
| Ti(IV)Pc-4CH$_3$ | -0.005 | 0.018 | 0.18 | $1.61E^{+11}$ | $2.16E^{+12}$ | | |
| Ti(IV)Pc-4C$_2$H$_5$ | 0.001 | -0.004 | 0.44 | $6.16E^{+8}$ | $7.26E^{+9}$ | | |

Moreover, for the concave molecular pair our calculated hole coupling is an order of magnitude higher and electronic coupling is an order of magnitude lower than that of reference 10. This may be due to the distance of the concave molecular pair. The crystallographic distance is reported as 3.326 Å [10] for concave system, however we used 3.118 Å convex pair distance, which is indeed shorter than the real distance of the molecules. Therefore, HOMOs, highly delocalized orbitals interact strongly to produce higher hole coupling. Therefore, corresponding hole transfer rates are three time higher than that of convex molecules. Nevertheless, the lower electronic coupling is not clear.

**Table 2** Reorganization energy calculated for Ti(II)Pc complexes.

| Complexes | Reorganization Energy ($\lambda$) eV |
|---|---|
| Ti(II)Pc | 0.27 |
| Ti(II)Pc-4F | 0.31 |
| Ti(II)Pc-4Cl | 0.09 |
| Ti(II)Pc-4I | 0.16 |
| Ti(II)Pc-4CH$_3$ | 0.28 |
| Ti(II)Pc-4C$_2$H$_5$ | 0.32 |

The Ti(IV)Pc unsubstitued molecule also has four orders of magnitude higher hole hopping rate (table-1). However, due to peripheral ligand substitution the hole migration rates reduced in Ti(IV)Pc. This reduction is three to four orders of magnitude for electron-withdrawing halogen ligands. Electron migration also reduced up to an order of magnitude in Iodine and fluorine substituted complexes. Nevertheless, electron hopping between chlorine substituted complexes do not affect significantly. In addition, chlorine substituted complex has similar electron and hole hopping rates. Thus it has higher electron hopping rate than that of other halogen substituted complexes. In Ti(IV)Pc halogen substituted complexes $\lambda$ values are very close to each other (see table-1). Whereas the coupling value for electrons (LUMO orbitals) is twice high in Ti(IV)Pc-4Cl complex. Thus the higher electron hopping rate of chlorine substituted complex is mainly attributed to the higher electronic coupling. Thus the higher electron hopping rate in chlorine Ti(IV)Pc-4Cl complex provide insight into higher charge transfer ability in this material. The Ti(IV)Pc halogen substituted complexes have shown higher reorganization energies with reference to their unsubstituted molecules. In fact, halogen substituted complexes have two times greater reorganization energies. On the contrary to that Ti(II)Pc halogen substituted complexes have two times lower reorganization energy values except for fluorine substituted complex. The Ti(II)Pc-4F complex has the similar reorganization energy value with respect to unsubstituted Ti(II)Pc. The Ti(II)Pc-4Cl complex has the lowest reorganization energy value reported among all the complexes that have been investigated in this study. That is 0.09 eV and lower than the reorganization energy value 0.16 eV of TiOPc. Therefore, we think that the similar high electron hopping mechanism may occur in [Ti(IV)Pc-4Cl]$^{2+}$ complex (either due to the higher electronic coupling or lower reorganization energy) and it facilitates electron transfer from the chlorine substituted titanyl-phthalocyanine complex to the adsorbate molecules. Thus, it reduced $H_2O_2$ spontaneously on the catalytic active site. However, lower electron hopping in iodine and fluorine substituted complexes reduced the efficiency of electron transfer towards adsorbate.

In contrast to electron-withdrawing ligand substitution, electron-donating ligand substituted complexes have higher electron hopping rates. On the other hand electron-donating substituents increased the reorganization energy in all the complexes except Ti(IV)Pc-4CH$_3$ complex. It has the lowest recorded $\lambda$ value for electron-donating substituents, which is 0.18 eV. This directly correlates with higher electron hopping rates of Ti(IV)Pc-4CH$_3$. In fact, the electron hopping rate is even better than that of TiOPc. The lower $\lambda$ values recorded in Ti(II)Pc electron-donating ligand complexes also provide indirect evidence for their higher charge mobility. Thus the reorganization energy has significant influence on the charge hopping between titanyl-phthalocyanine molecules.

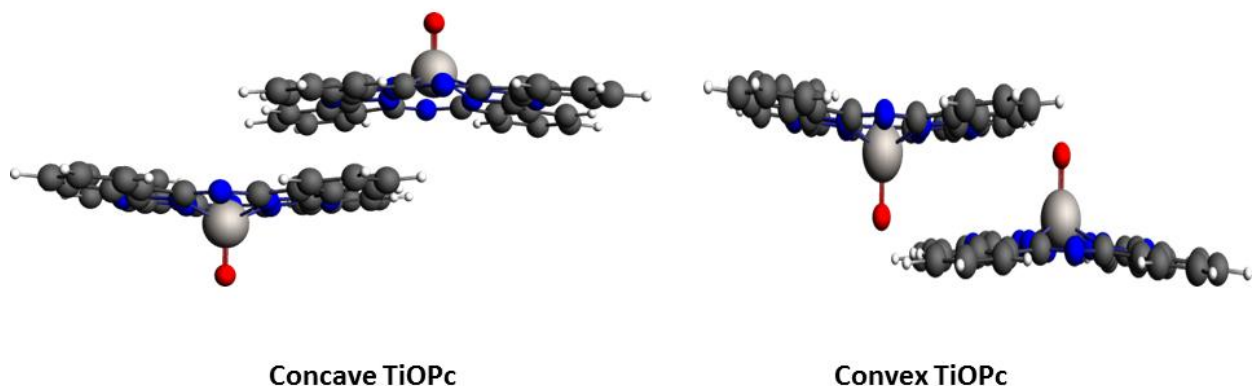

**Fig- 4** The concave and convex arrangements of TiOPc molecule. Carbon grey, hydrogen white, nitrogen blue, titanium silver and oxygen is red in color.

## Conclusions

Here we investigated electronic coupling, reorganization energies and Markus hopping rates of known and unknown titanyl-phthalocyanine systems. Higher charge mobility and reorganization energy associated with electron transfer of TiOPc are predicted with reasonable accuracy. Therefore, our calculations are in good agreement with available literature. Low reorganization energy values in Ti(II)Pc complexes provide explanation for the origin of higher charge transfer ability. In addition, Ti(IV)Pc chlorine substituted complex has higher electron hopping rate due to higher electronic coupling. Therefore, MPcs with higher electron hopping rates can be speculated as potential catalysts for ORR. The electron donating substituents increased the electron hopping rates of the Ti(IV)Pc complexes. Therefore, substituting electron-donating ligands may stimulate the interaction between LUMO orbitals of the complexes and increase the electron mobility. Therefore, electron-donating ligand substitution may be used to increase the electron mobility in n-channel organic field-effect transistors.

**Acknowledgements**

We gratefully acknowledge PhD scholarship from Chinese Scholarship Council (CSC). JRD thanks to Jacob's university of Bremen for the computational facility provided for this work.